\documentclass{article}

\usepackage{amsfonts,amssymb,epsfig,amsthm}

\newcommand{\R}{\mathbb{R}}
\newcommand{\C}{\mathbb{C}}
\newcommand{\CC}{\mathcal{C}}
\newcommand{\I}{\mathrm{i}}
\newcommand{\E}{\mathrm{e}}
\newcommand{\D}{\mathrm{d}}
\newcommand{\N}{\mathbb{N}}

\newcommand{\T}{\mathbb{T}}

\newcommand{\B}{\mathcal{B}}

\newcommand{\A}{\mathcal{A}}
\newcommand{\Or}{\mathcal{O}}
\newcommand{\Lap}{\Delta}
\newcommand{\epsi}{\varepsilon}

\newcommand{\vp}{\times}
\newcommand{\M}{\mathcal{M}}

\newcommand{\be}{\begin{equation}}
\newcommand{\ee}{\end{equation}}
\newcommand{\er}{\hfill$\diamondsuit$}

\newtheorem{theorem}{Theorem}

\newtheorem{proposition}[theorem]{Proposition}
\newtheorem{corollary}[theorem]{Corollary}
\newtheorem*{assumption}{Assumption}
\theoremstyle{definition}
\newtheorem*{definition}{Definition}
\newtheorem*{remark}{Remark}
\newtheorem*{notation}{Notation}

\begin{document}
\pagenumbering{arabic}

\title{Propagation of Wigner  functions for the Schr\"odinger equation
with  a perturbed periodic potential}

\author{Stefan Teufel and Gianluca Panati\\[2mm]Zentrum Mathematik, TU M\"unchen, Germany\\  panati@ma.tum.de, teufel@ma.tum.de}
\date{June 30, 2003}

\maketitle
\begin{abstract}
Let $V_\Gamma$ be a lattice periodic potential and $A$ and
$\phi$  external electromagnetic potentials which vary slowly on the scale set by the lattice spacing.
It is shown that the Wigner function of a solution of the Schr\"odin\-ger equation
with Hamiltonian operator
$H = \frac{1}{2} ( -\I\nabla_x - A(\varepsilon x) )^2 + V_\Gamma (x) + \phi(\varepsilon x)$
propagates along the flow of the semiclassical model of solid states physics up an error of order $\epsi$.
If $\epsi$-dependent corrections to the flow are taken into account, the error is improved to order $\epsi^2$.
We also discuss the propagation of the Wigner measure.
The results are obtained as corollaries of an Egorov type theorem proved in \cite{PST3}.
\end{abstract}

\markboth{Stefan Teufel and Gianluca Panati}{Effective dynamics in periodic potentials}

\section{Introduction}

One of the central questions of solid state physics is to understand the
motion of electrons in the periodic potential which is generated by
the ionic cores. While this problem is quantum mechanical, many
electronic properties of solids can be understood already in the
  semiclassical approximation
\cite{AsMe,Ko,Za}. One argues that for suitable wave packets,
which are spread over many lattice spacings, the main effect of a
periodic potential   $V_{\Gamma}$ on the
electron dynamics corresponds to changing the dispersion relation
from the free kinetic energy $E_{\rm free}(p) = \frac{1}{2}\, p^2$
to the modified kinetic energy $E_n(p)$ given by the $n^{\rm th}$
Bloch function. Otherwise the electron responds to slowly varying external
potentials $A$, $\phi$ as in the case of a vanishing periodic
potential. Thus the semiclassical equations of motion are
\be  \label{semiclassical_dynamics}
\dot r = \nabla  E_n(\kappa ) \,,\qquad \dot \kappa  = -\nabla \phi(r) +
\dot r \vp B(r)\,,
\ee
where $\kappa = k - A(r)$ is the kinetic momentum and $B= \mathrm{curl}
A$ is the magnetic field. (We choose units in which
the Planck constant $\hbar$,   the speed $c$ of light, and
the mass $m$   of the electron
 are equal to one, and absorb the charge $e$ into the potentials.)
The corresponding equations of motion
for the canonical variables $(r,k)$ are generated by the
Hamiltonian
\be\label{Hsc}
H_{\rm sc} (r,k) = E_n\big(k-A(r)\big) + \phi(r)\,,
\ee
where $r$ is the position and $k$ the quasi-momentum of the
electron. Note that there is a semiclassical evolution for each
Bloch band separately. The distinction between the canonical variable $k$,
the Bloch- or quasi-momentum, and the kinetic momentum $\kappa = k-A(r)$
is often not made explicit in the physics literature. It is, however, crucial
for the formulation of the precise connection between the semiclassical equations of motion
(\ref{semiclassical_dynamics}) and the underlying Schr\"odinger equation (\ref{scaled_dynamics}).

In \cite{PST3} we use adiabatic perturbation theory in order to
understand on a mathematical level how these semiclassical
equations emerge from the underlying Schr\"odinger equation
\be \label{basic_dynamics}
\I\,  \partial_s\,\psi(y,s)= \left(
{\textstyle\frac{1}{2}}\big(-\I \nabla_y - A(\epsi y)\big)^2 +
V_{\Gamma}(y) + \phi(\epsi y)\right)
 \psi(y,s)
\ee
in the limit $\epsi\to 0$ at leading order. In addition, the order
$\epsi$  correction to (\ref{semiclassical_dynamics}) are established, see Equation (\ref{Semi1}).

In (\ref{basic_dynamics}) the potential  $V_{\Gamma}:\R^d\to \R$ is
periodic with respect to some regular lattice $\Gamma$ generated
through the basis $\{\gamma_1,\ldots,\gamma_d\}$,
$\gamma_j\in\R^d$, i.e.\
\[
\Gamma =\Big\{ x\in\R^d: x= \textstyle{\sum_{j=1}^d}\alpha_j\,\gamma_j
\,\,\,\mbox{for some}\,\,\alpha \in \mathbb{Z}^d \Big\}
\]
and $V_{\Gamma}(\,\cdot + \gamma) = V_{\Gamma}(\cdot)$ for all
$\gamma\in\Gamma$. The lattice spacing defines the microscopic
spatial scale. The external potentials $A(\epsi y)$ and $\phi(\epsi
y)$, with $A:\R^d\to\R^d$ and $\phi:\R^d\to\R$, are slowly varying on
the scale of the lattice, as expressed through the  dimensionless
scale parameter $\epsi$, $\epsi\ll 1$. In particular, this means
that the external fields are weak compared to the fields generated
by the ionic cores, a condition which is
 satisfied for real metals  even for the strongest external electrostatic fields
available and for a wide range of  magnetic fields, cf.\
\cite{AsMe}, Chapter~12.

Note that the external forces due to $A$ and  $\phi$ are of order
$\epsi$ and therefore have to act over a time of order
$\epsi^{-1}$ to produce finite changes, which is taken as the definition of
the macroscopic time scale. Hence, one is interested in solutions of (\ref{basic_dynamics})
for macroscopic times. The macroscopic space-time scale
$(x ,t)$ is defined through $x =\epsi y$ and $t=\epsi s$. With
this change of variables  Equation  (\ref{basic_dynamics}) reads
\be  \label{scaled_dynamics}
\I\,\epsi\, \partial_{t }\,\psi^\epsi(x ,t )= \left( {\textstyle
\frac{1}{2}}\big(-\I \epsi\nabla_{x } - A(x )\big)^2 +
V_{\Gamma}(x /\epsi) + \phi(x)\right)
 \psi^\epsi(x ,t )
\ee
with initial conditions
$\psi^\epsi(x )=\epsi^{-d/2}\psi(x /\epsi)$. If $V_{\Gamma}=0$, then the limit $\epsi\to 0$ in
Equation (\ref{scaled_dynamics}) is the usual semiclassical limit
with $\epsi$ replacing $\hbar$.

The problem of deriving (\ref{semiclassical_dynamics})  from
the Schr\"odinger equation
(\ref{basic_dynamics}) in the limit $\epsi\to 0$ has been attacked
along several routes. In the physics literature
(\ref{semiclassical_dynamics}) is usually accounted for  by
constructing suitable semiclassical wave packets. We refer to
\cite{Lu,Ko,Za}. The few mathe\-matical approaches  to the
time-dependent problem   (\ref{scaled_dynamics}) extend
techniques from semiclassical analysis, as the Gaussian beam construction
\cite{GRT,DGR}, or   Wigner measures \cite{GMMP,BFPR,BMP}.

In this note we explain and elaborate on recent results from \cite{PST3}.
In \cite{PST3} we derived (\ref{semiclassical_dynamics}) from (\ref{scaled_dynamics})
for quite general external potentials $A$ and $\phi$. The construction is based on
the space-adiabatic perturbation theory developed in  \cite{PST1,Te}, see also \cite{NeSo} and the contribution of G.~Nenciu in the present volume.
The crucial observation is that the step from (\ref{basic_dynamics}) to
(\ref{semiclassical_dynamics}) involves actually two approximations.
Semiclassical behavior can only emerge if a Bloch band is
separated by a gap from the other bands and thus the corresponding
subspace decouples adiabatically from its orthogonal complement.
The dynamics inside this adiabatic subspace is governed by an effective Hamiltonian $\widehat h_{\rm eff}^\epsi$,
which is explicitly given as an $\epsi$-pseudodifferential operator.  Eventually, the semiclassical limit of
$\widehat h_{\rm eff}^\epsi$  leads to (\ref{semiclassical_dynamics}).

Hence   (\ref{basic_dynamics}) needs to be reformulated as a
space-adiabatic problem. This has been done first in \cite{HST} for the case of zero
 magnetic field and then in \cite{PST3} for general electric and magnetic fields.
  The  results   obtained in this way
constitute not only the derivation of the semiclassical model
(\ref{semiclassical_dynamics}) in this generality, but   they
allow to compute systematically higher order corrections in the
small parameter $\epsi$. It turns out that the electron acquires a
$k$-dependent electric moment $\mathcal{A}_n(k)$ and magnetic
moment $\mathcal{M}_n(k)$. If the $n^{\rm th}$ band is
nondegenerate (and isolated) with Bloch eigenfunctions
$\psi_n(k,x)$, the electric dipole moment is given by the Berry
connection \be \A_{n} (k) = \I \,\big\langle \psi_{n}(k), \nabla
\psi_{n}(k) \big\rangle \,. \ee and the magnetic moment by the
Rammal-Wilkinson phase \be \mathcal{M}_{n}(k) =
\textstyle{\frac{\I}{2}}\,\big\langle \nabla\psi_{n}(k), \ \vp
(H_{\rm per}(k) - E(k))\nabla\psi_{n}(k) \big\rangle\,. \ee
 Here $\langle \cdot \, , \cdot \rangle \, $ is the inner
product in $L^2(\R^d/\Gamma)$ and $H_{\rm per}(k)$ is $H$ of
(\ref{basic_dynamics}) with $\phi = 0 = A$ for fixed Bloch
momentum $k$.
Note that $E_n$, $\A_n$ and $\M_n$ are $\Gamma^*$-periodic functions of $k$,
where  $\Gamma^*$ is the lattice dual to $\Gamma$. Hence one can as well think of them as functions
on the domain $M^* = \R^d/\Gamma^*$, the first Brillouin zone.

The   semiclassical equations of motion including first order corrections  read
\begin{eqnarray}
\dot r &=& \nabla_{\kappa} \Big( E_n(\kappa) - \epsi \, B(r)\cdot \M_n(\kappa
)\Big)
- \epsi\, \dot  \kappa \vp \Omega_n(\kappa)\,,\nonumber\\\label{Semi1}\\
\dot  \kappa  &=&  -\nabla_r \Big(\phi(r) - \epsi\, B(r)\cdot
\M_n(\kappa)\Big) +\dot r \vp B(r) \,.\nonumber
\end{eqnarray}
with $\Omega_n(k) = \nabla \vp \A_n(k)$ the curvature of the Berry
connection.

In order to state the precise connection between the semiclassical equations of motion
(\ref{semiclassical_dynamics}) resp.\ their refined version (\ref{Semi1})
and the underlying Schr\"odinger equation (\ref{scaled_dynamics}), we need some more notation.
Let
\be\label{6b}
H^\epsi =
  {\textstyle
\frac{1}{2}}\big(-\I \epsi\nabla_{x } - A(x )\big)^2 +
V_{\Gamma}(x /\epsi) + \phi(x)
\ee
be the Hamiltonian of (\ref{scaled_dynamics}). Under
the following assumption on the potentials, which  will be imposed throughout,
$H^\epsi$ is self-adjoint on $H^2(\R^d)$. Here $C^\infty_{\rm b}(\R^d)$ denotes the space of smooth functions
which are bounded together with all their derivatives.
\begin{assumption} Let $V_\Gamma$ be
infinitesimally bounded with respect to $-\Lap$ and assume that $\phi \in
C^\infty_{\rm b}(\R^d,\R)$ and $A_j\in C^\infty_{\rm b}(\R^d,\R)$
 for
any  $j\in\{ 1,\ldots,d \} $.
\end{assumption}
To each isolated Bloch band $E_n$ there corresponds  an associated almost invariant band-subspace $\Pi^\epsi_n L^2(\R^d)$.
The orthogonal projector $\Pi^\epsi_n$ onto this subspace is constructed in \cite{PST3}.
Only  for states which start in this subspaces and thus, by construction, remain there
up to small errors,    the semiclassical equations of motion (\ref{Semi1})
can have any significance.

The flow of the dynamical system (\ref{Semi1}) is denoted by
$\Phi^t_\epsi:\R^{2d}\to\R^{2d}$ or in canonical coordinates $(r,k)=(r,\kappa +A(r))$ by
 \[
\overline \Phi^t_\epsi (r,k) = \Big(
\Phi^t_{\epsi\,r}\big(r,k-A(r)\big),\,\Phi^t_{\epsi\,\kappa}\big(r,k-A(r)\big)+A(r)\Big)\,.
\]
The existence of the smooth family of diffeomorphisms $\Phi^t_\epsi$ is not completely
obvious from   (\ref{Semi1}) alone, but follows from the Hamiltonian formulation of (\ref{Semi1})
presented in the next section.

\begin{notation} Throughout this paper we will use the Fr\'echet space
\[
\CC  = C^\infty_{\rm b}(\R^{2d})\,,
\]
equipped with the metric $d_\CC$ induced by the standard family of semi-norms
\[
\|a\|_\alpha = \|\partial^\alpha a\|_\infty\,,\quad \alpha\in \N^{2d}_0\,,
\]
and the subspace
of $\Gamma^*$-periodic observables
\[
\CC_{\rm per} = \{ a\in \CC: a(r,k+\gamma^*)= a(r,k) \,\,\forall\, \gamma^*\in\Gamma^*\} \,.
\]
We abbreviate $d_\CC(a):= d_\CC(a,0)$.\er
\end{notation}

The main result of \cite{PST3} on the semiclassical
limit of (\ref{scaled_dynamics}) is the following Egorov type theorem.

\begin{theorem} \label{EgCor}
Let $E_n$ be an isolated, non-degenerate Bloch band.
For each finite time-interval $I\subset \R$ there is a
constant $C<\infty$, such that for all $a\in \CC_{\rm per}$ with Weyl quantization  $\widehat a = a(  x,-\I\epsi\nabla_x)$ one has
\be\label{n}
 \left\|\,   \left(\, \E^{\I H^\epsi t/\epsi} \, \widehat a\,\,  \E^{-\I
  H^\epsi t/\epsi}\,-\, \widehat{ a\circ \overline \Phi^{t}_0 }\, \right)\,\Pi^\epsi_n \,\right\|_{\B(L^2(\R^d))}
  \leq \epsi\,C\,d_\CC(a)
\ee
and
\be\label{n2}
 \left\|\, \Pi^\epsi_n\, \left(\, \E^{\I H^\epsi t/\epsi} \, \widehat a\,\,  \E^{-\I
  H^\epsi t/\epsi}\,-\, \widehat{ a\circ \overline \Phi^{t}_\epsi }\, \right)\,\Pi^\epsi_n \,\right\|_{\B(L^2(\R^d))}
  \leq \epsi^2\,C\,d_\CC(a)\,.
\ee
\end{theorem}

\begin{remark}
The corresponding statement in \cite{PST3} does not make explicit the dependence of the error on the observable $a$.
However, the more precise version formulated here is a standard consequence of the Calderon-Vaillancourt theorem
and the fact that composition with $\overline \Phi^t_\epsi$ is a continuous map from $\CC$ into itself.\er
\end{remark}

\begin{remark}
On an abstract level the distinction between the functions $\Phi^t_\epsi$ and $\overline \Phi^t_\epsi$ is immaterial,
since both functions express the same dynamical flow in two systems of coordinates.
However, the distinction between the systems of coordinates becomes important when the quantization is considered.
The Weyl quantization appearing in (\ref{n}) and (\ref{n2}) must be understood with respect to the system
of coordinates $(r,k)$. Analogous consideration hold true for formulas involving a Wigner transform, as in Corollary~\ref{WigCor}.\er
\end{remark}

The main objective of this note is to elaborate on Theorem~\ref{EgCor} in order to make contact to
alternative approaches and results on the semiclassical limit of (\ref{scaled_dynamics}).
This are, as mentioned above,   Wigner functions \cite{GMMP,BFPR,BMP},
semiclassical wave packets \cite{Lu,Ko,Za,SuNi} and WKB-type solutions of
(\ref{scaled_dynamics}) \cite{Bu,GRT,DGR}. We focus on the semiclassical transport of
 Wigner functions and Wigner measures in the following.
 Before we do so,
it is worthwhile to first examine the equations of motion
(\ref{Semi1}) in some more detail.

\section{The refined semiclassical equations of motion}

The dynamical equations (\ref{Semi1}), which define the
$\epsi$-corrected semiclassical model, can be written as
\begin{equation} \label{SC model}
\begin{array}{ccc}
    \dot r & = & \nabla_\kappa H_{\rm sc}(r,\kappa)  -   \epsi \, \dot \kappa \times \Omega_n(\kappa)\,, \\
    &&\\
    \dot \kappa & = & \hspace{-3mm}-  \nabla_r H_{\rm sc}(r,\kappa) +   \dot r \times B(r)
\end{array}
\end{equation}
with \be \label{SC energy} H_{\rm sc}(r,\kappa) := E_n(\kappa) +
\phi(r) - \epsi \, \mathcal{M}_n(\kappa) \cdot B(r)\,. \ee We
shall show that (\ref{SC model}) are the Hamiltonian equations of
motion for (\ref{SC energy}) with respect to a suitable
$\epsi$-dependent symplectic form $\Theta_{B,\epsi}$. The
semiclassical equations of motion (\ref{Semi1}) are defined for
arbitrary dimension $d$. However, to simplify presentation, we use
a notation motivated by the vector product and the duality between
1-forms and 2-forms for $d=3$, which we briefly explain.

\begin{notation}
If $d\not= 3$, then $B$, $\Omega_n$ and $\M_n$ are 2-forms with components
\[
B_{ij}(r) = \partial_i A_j(r) -\partial_j A_i(r) \,,\qquad \Omega_{ij}(k) = \partial_i \A_j(k) -\partial_j \A_i(k)
\]
and
\[
\M_{ij}(k) = {\rm Re} \,  {\textstyle \frac{\I}{2}} \,  \big\langle \partial_i\psi_n(  k),
  (H_{\rm per} - E)( k) \
\partial_j \psi_n (  k) \big\rangle \,.
\]
For $d=3$ a 2-form $B_{ij}(r)$ is naturally associated with the vector $B_k(r) = \epsilon_{kij} \,B_{ij}(r)$.
We use the convention that summation over repeated indices is implicit.
Then in (\ref{Semi1}) the inner product $B\cdot \M_n$ refers to the product of the associated vectors and we generalize
the notation to arbitrary dimension $d$ using
the inner product of 2-forms defined through
\[
B \cdot \M :=  *^{-1}(B \wedge *\M)    =\sum_{j=1}^d \sum_{i=1}^d  B_{ij} \M_{ij}\,,
\]
where $*$ denotes the Hodge duality induced by the euclidian metric.
In the same spirit for a vector field $w$ and a 2-form $F$ the generalized ``vector product'' is
\[
(w \vp F)_j := (*^{-1}(w \wedge *F))_j = \sum_{i=1}^d  w_i
F_{ij}\,,
\]
where the duality between 1-forms and vector fields is used implicitly. \er
\end{notation}

We keep fixed the system of coordinates $z=(r,\kappa)$ in $\R^{2d}$ for the following. The
standard symplectic form $\Theta_0 = \Theta_{0}(z)_{lm}\, \D z_m
\wedge \D z_l $, where $l,m \in\{ 1,\ldots ,2d\}$, has
coefficients given by the constant matrix
\[ \label{Symplectic standard} \Theta_{0}(z)= \left(
\begin{array}{cc}
  0 & -\mathbb{I} \\
  \mathbb{I} & 0
\end{array}\right)\,,
\]
where $\mathbb{I}$ is the identity matrix in $\mathrm{Mat}(d,\R)$.
The symplectic form, which turns (\ref{SC model}) into Hamilton's
equation of motion for $H_{\rm sc}$, is given by the  2-form
$\Theta_{B,\, \epsi}= \Theta_{B,\,\epsi}(z)_{lm}\,  \D z_m \wedge
\D z_l$ with coefficients \be \label{Symplectic matrix}
\Theta_{B,\, \epsi}(r,\kappa)= \left(
\begin{array}{cc}
  B(r) & -\mathbb{I} \\
  \mathbb{I} &  \epsi \ \Omega_n(\kappa)
\end{array}\right)\,.
\ee For $\epsi=0$ the 2-form $\Theta_{B,\, \epsi}$ coincides with
the magnetic symplectic form $\Theta_{B}$ usually employed to
describe in a gauge-invariant way  the motion of a particle in a
magnetic field (\cite{MaRa}, Section~6.6). For $\epsi$ small
enough, the matrix (\ref{Symplectic matrix})   defines a
symplectic form, i.e.\ a closed non-degenerate 2-form.

 With these definitions the corresponding Hamiltonian equations are
\[ \Theta_{B,\, \epsi}(z) \ \dot{z} = \D H_{\rm sc}(z)\,,
\]
or equivalently
\[
\left(
\begin{array}{cc}
  B(r) & -\mathbb{I} \\
  \mathbb{I} & \epsi \ \Omega_n(\kappa)
\end{array}\right)
\left(
\begin{array}{c}
  \dot r \\
  \dot \kappa
\end{array}\right) =
\left(
\begin{array}{c}
  \nabla_r H(r,\kappa) \\
  \nabla_\kappa H(r,\kappa)
\end{array}\right)\,,
\]
which agrees with (\ref{SC model}). We notice that this discussion
remains valid if  $\Omega_n$ admits a potential only locally, as
it happens generically for magnetic Bloch bands.

 The symplectic structure is therefore determined by the magnetic field $B(r)$
 and by the curvature of the Berry connection $\Omega(k)$, which encodes relevant information
 about the geometry of the Bloch bundle $\psi_n(k,\cdot) \mapsto k\in M^*$.
 One can show  that, whenever the Hamiltonian $H_{\rm per}$ has time-reversal symmetry
  one has that $\Omega_n(-k) = -\Omega_n(k)$. Moreover,
 if the lattice $\Gamma$ has a center of inversion, then  $\Omega_n(-k) = \Omega_n(k)$.
 Thus, the two symmetries together imply that $\Omega_n(k)$ vanishes pointwise.
 But there are many crystals which do {\em not} have a center of inversion and, more important,
 in the presence of a strong uniform magnetic field the time-reversal symmetry is broken.
 The latter is the typical setup to describe the Quantum Hall Effect, a situation  in which the curvature of the
 Berry connection plays a prominent
 role.
 Indeed,
the equations of motion (\ref{Semi1})
provide a simple semiclassical explanation of the Quantum Hall
Effect. Let us specialize (\ref{Semi1}) to two dimensions and take $B(r)=0$,
$\phi(r) = - \mathcal{E} \cdot r$, i.e.\ a weak driving electric
field and a strong uniform magnetic field with rational flux. Then, since $\kappa=k$,
the equations of motion   become $\dot r = \nabla_k
E_n(k) + \mathcal{E}^\perp\Omega_n(k)$, $\dot k = \mathcal{E}$,
where $\Omega_n$ is now scalar, and $\mathcal{E}^\perp$ is
$\mathcal{E}$ rotated by $\pi/2$. We assume initially $k(0)=k$ and
a completely filled band, which means to integrate with respect to
$k$ over the first Brillouin zone $M^*$. Then the average current for
band $n$ is given by
\[
j_n = \int_{M^{*}} \D k \,\dot r(k) = \int_{M^{*}} \D k
\,\big(\nabla_{ k } E_n( k )-     \mathcal{E}^\perp \Omega_n( k
)\big) =  - \mathcal{E}^\perp \int_{M^{*}} \D k\,\Omega_n( k)\,.
\]
$\int_{M^{*}} \D k\,\Omega_n( k)$ is the Chern number of the
magnetic Bloch bundle and as such an integer, cf.\ \cite{TKNN}. Further applications
related to the semiclassical first order corrections are the anomalous
Hall effect \cite{JNM} and the thermodynamics of the Hofstadter
model \cite{GaAv}.

\section{Semiclassical transport of Wigner functions}

Theorem~\ref{EgCor} provides a semiclassical description of the evolution of
 observables. The most direct way to turn it into a description
for the semiclassical evolution of states is via duality, i.e.\ via the Wigner function.
Recall that according to the Calderon-Vaillancourt theorem there is a constant $C<\infty$ depending only on the
dimension $d$ such that for $a\in\CC$ one has
\be\label{CV}
|\,\langle \psi, \,\widehat a \,\psi\rangle_{L^2(\R^d)} \,| \leq \,C\,d_\CC(a)\,\|\psi\|^2\,.
\ee
Hence, the map $\CC\ni a\mapsto \langle \psi, \,\widehat a \,\psi\rangle\in \C$ is continuous
and thus defines an element $w^\psi_\epsi$ of the dual space $\CC'$, the Wigner function of $\psi$.
Writing
\be\label{Duality}
\langle \psi, \,\widehat a \,\psi\rangle =:
\langle w^\psi_\epsi, \,a\rangle_{\CC',\CC}
 =:\int_{\R^{2d}}\D q\,\D p \, \,
 a(q,p)\,w^\psi_\epsi(q,p)
\ee
and inserting into (\ref{Duality}) the definition of the Weyl quantization for $a\in \mathcal{S}(\R^{2d})$
\[
 ( \widehat a \psi  ) (x)=\frac{1}{(2\pi  )^d}
\int_{\R^{2d}} \D\xi\, \D y\,
a\big(\textstyle{\frac{1}{2}}(x+y),\epsi \xi \big)\, \E^{\I \xi \cdot
(x-y) }\, \psi (y)\,,
\]
one arrives at the formula
\be\label{Wigner}
w^\psi_\epsi(q,p)  =\frac{1}{(2\pi  )^d} \int_{\R^{d}}\,\D\xi\,
\E^{\I\xi\cdot p}\,
\psi^*(q+\epsi\xi/2)\,\psi (q-\epsi\xi/2)
\ee
for the Wigner function.
Direct computation yields
 \[
 \|\,w^\psi_\epsi \,\|_{L^2(\R^{2d})} = \epsi^{-d}\,(2\pi)^{-d/2}\,\|\,\psi\,\|^2_{L^2(\R^d)}\,.
 \]
Therefore, $w^\psi_\epsi\in L^2(\R^{2d})$ for all $\epsi>0$, which explains  the notion of
Wigner function.  Although $w^\psi_\epsi$ is obviously real-valued, it attains also negative values in general.
Hence, it does not define a probability distribution on phase space.  However, it correctly produces
quantum mechanical distributions via (\ref{Duality}).

With this preparations we obtain the following corollary of Theorem~\ref{EgCor},
which says that the Wigner function of the solution of the Schr\"odinger equation
(\ref{scaled_dynamics}) is approximately transported along the classical flow of (\ref{semiclassical_dynamics}) resp.\
(\ref{Semi1}).

\begin{corollary}\label{WigCor}
Let $E_n$ be an isolated, non-degenerate Bloch band.
 Then for each finite time-interval $I\subset \R$ there is a
constant $C<\infty$ such that for $t\in I$,  $a\in\CC_{\rm per}$ and for   $\psi_0\in \Pi^\epsi_n L^2(\R^d)$ one has
\[
 \left|  \left\langle \left( w^{\psi_t}_\epsi- w^{\psi_0}_\epsi\circ \overline\Phi_0^{\,-t}\right) , a \right\rangle_{\CC',\CC}   \right|\leq \epsi\,C\,d_\CC(a)\,\|\psi_0\|^2
\]
and
\[
 \left|  \left\langle \left( w^{\psi_t}_\epsi- w^{\psi_0}_\epsi\circ \overline\Phi_\epsi^{\,-t}\right) , a \right\rangle_{\CC',\CC}   \right|\leq \epsi^2\,C\,d_\CC(a)\,\|\psi_0\|^2\,.
\]
Here $\psi_t = \E^{-\I H^\epsi t/\epsi}\psi_0$ is the solution of the Schr\"odinger equation (\ref{scaled_dynamics}).
\end{corollary}

\begin{remark}
When proving results for the transport of Wigner functions or Wigner measures it is common, e.g.\  \cite{GMMP,MMP,BMP},
to write down the transport equation for $w_\epsi(t) := w^{\psi_0}_\epsi\circ \overline\Phi_0^{\,-t}$ instead of
using the flow  $\overline\Phi_0^{t}$. Clearly our results can be reformulated in this way, cf.\ Corollary~\ref{MeasureCor},
but the resulting transport equation looks complicated compared to the simple dynamical system (\ref{semiclassical_dynamics})
governing  its characteristics.
\er\end{remark}

\begin{proof}[Proof of Corollary \ref{WigCor}]
The result is rather a reformulation of Theorem~\ref{EgCor} than a real corollary.
According to the Definition  (\ref{Duality}) and  Theorem~\ref{EgCor} one has
\begin{eqnarray*}
  \langle   w^{\psi_t}_\epsi  , a  \rangle_{\CC',\CC} &=&
 \langle \psi_t, \,\widehat a\,\psi_t\rangle_{L^2(\R^d)} \\&=&
 \langle \psi_0, \,   \E^{\I H^\epsi t/\epsi} \, \widehat a\,\,  \E^{-\I
  H^\epsi t/\epsi}\, \psi_0 \rangle_{L^2(\R^d)}
 \\&=& \langle \psi_0, \,\Pi^\epsi_n\,   \E^{\I H^\epsi t/\epsi} \, \widehat a\,\,  \E^{-\I
  H^\epsi t/\epsi}\, \Pi^\epsi_n\,\psi_0 \rangle_{L^2(\R^d)} \\
  &=& \langle \psi_0,\, \Pi^\epsi_n\,   \widehat{ a\circ \overline  \Phi^{t}_\epsi }\, \Pi^\epsi_n\,\psi_0\rangle_{L^2(\R^d)} +\Or(\epsi^2) \\
  &=& \langle \psi_0,\,   \widehat{ a\circ \overline  \Phi^{t}_\epsi }\,  \psi_0\rangle_{L^2(\R^d)} +\Or(\epsi^2)\,.
 \end{eqnarray*}
Since the map $\CC\ni a\mapsto a\circ \overline \Phi^{t}_\epsi\in \CC$ is continuous, the duality relation (\ref{Duality}) can be applied again and yields
 \[
  \langle \psi_0,\,    \widehat{ a\circ \overline \Phi^{t}_\epsi }  \,\psi_0\rangle_{L^2(\R^d)}
  =   \left\langle w^{\psi_0}_\epsi, a\circ \overline\Phi_\epsi^{\,t}\right\rangle_{\CC',\CC}
  =  \left\langle w^{\psi_0}_\epsi\circ \overline\Phi_\epsi^{\,-t}, a\right\rangle_{\CC',\CC}
   \,.
\]
\end{proof}

Since the functions $E_n$, $\M_n$ and $\Omega_n$ appearing
in the equations of motion (\ref{Semi1}) are all $\Gamma^*$ periodic,
the natural phase space for the flow (\ref{Semi1}) is $\R^d\times \T^*$ rather than $\R^{2d}$. Here
$\T^d := \R^d/\Gamma^*$ is the first Brillouin zone $M^*$ equipped with periodic boundary conditions.
Hence one can fold  the Wigner transform onto the first Brillouin zone and define
\be
w^\psi_{\epsi\,\rm red}(r,k) = \sum_{\gamma^*\in \Gamma^*} w^\psi_\epsi (r,k+\gamma^*)\quad\mbox{for}\quad (r,k)\in \R^d\times \T^d\,.
\ee
Then for periodic observables $a$ it follows that
\begin{eqnarray*}
\int_{\R^{2d}}\D r\,\D k\,\, a(r,k) \,   w^\psi_\epsi (r,k)
& =&   \sum_{\gamma^*\in\Gamma^*}  \int\limits_{\R^{d}\times M^*}\hspace{-6pt}\D r\,\D k\,\, a(r,k+\gamma^*) \,   w^\psi_\epsi (r,k+\gamma^*)\\
& =& \sum_{\gamma^*\in\Gamma^*} \int\limits_{\R^{d}\times M^*}\hspace{-6pt}\D r\,\D k\,\, a(r,k ) \,   w^\psi_\epsi (r,k+\gamma^*)\\
&=&  \int\limits_{\R^{d}\times \T^d }\hspace{-6pt}\D r\,\D k\,\, a(r,k) \,   w^\psi_{\epsi\,\rm red} (r,k)\,.
\end{eqnarray*}
Thus the statement of Corollary~\ref{WigCor} in terms of the reduced Wigner function becomes
  \[
   \langle \psi_t, \,\widehat a\,\psi_t\rangle_{L^2(\R^d)} = \int_{\R^{d}\times\T^d }\D r\,\D k\,\, a(r,k) \, \left( w^{\psi_0}_{\epsi\,\rm red}\circ \overline\Phi_\epsi^{\,-t}\right)
 (r,k) +\Or(\epsi^2)\,.
\]
Note that the reduced Wigner function $w^\psi_{\epsi\,\rm red}$ coincides with the ``band-Wigner function''  of
\cite{MMP} and the ``Wigner series'' of \cite{BMP}, both defined as
\[
w^\psi_{\epsi\,\rm s} (r,k) =  \frac{1}{|M^*|} \sum_{\gamma\in\Gamma} \,
\E^{\I\gamma\cdot k}\,  \psi(r+\epsi\gamma/2)\,\psi^* (r-\epsi\gamma/2)\,.
\]
This follows by a simple computation on the dense set $\psi\in \mathcal{S}(\R^d)$:
\begin{eqnarray*}
w^\psi_{\epsi\,\rm red}(r,k) &=& \sum_{\gamma^*\in \Gamma^*} w^\psi_\epsi (r,k+\gamma^*)\\
&=& \frac{1}{(2\pi)^d} \sum_{\gamma^*\in \Gamma^*} \int_{\R^{d}}\,\D\xi\,
\E^{\I\xi\cdot \gamma^*}\,\E^{\I\xi\cdot k}\,
\psi(r+\epsi\xi/2)\,\psi^* (r-\epsi\xi/2)\\
&=&  \frac{1}{|M^*|} \int_{\R^{d}}\,\D\xi\, \delta_\Gamma(\xi)\,
\E^{\I\xi\cdot k}\,
\psi(r+\epsi\xi/2)\,\psi^* (r-\epsi\xi/2)\\
&=&  \frac{1}{|M^*|} \sum_{\gamma\in\Gamma} \,
\E^{\I\gamma\cdot k}\,  \psi(r+\epsi\gamma/2)\,\psi^* (r-\epsi\gamma/2)\,,
\end{eqnarray*}
where $\delta_\Gamma(\xi)= \sum_{\gamma\in \Gamma} \delta(\xi-\gamma)$. We used the Poisson formula
\[
\frac{1}{(2\pi)^d}\sum_{\gamma^*\in \Gamma^*} \E^{\I\xi\cdot \gamma^*} = \frac{1}{|M^*|}\delta_\Gamma(\xi)\,.
\]

\section{Classical transport of the Wigner measure}

We now turn to the Wigner measure. Recall that
the Wigner function $w^\psi_\epsi(q,p)$ can be negative and, as a consequence, does not define a probability
distribution on phase space. In the limit $\epsi\to 0$ however,  $w^\psi_\epsi$  weakly converges to
a positive finite Radon measure $\mu^\psi\in \mathcal{M}^+_{\rm b}(\R^{2d})$ on phase space $\R^{2d}$, the Wigner measure of $\psi$.
For surveys on Wigner measures see e.g.\ \cite{LiPa,GMMP}.

\begin{proposition} \label{WigMeasure}
 Let $\epsi_j\stackrel{j\to \infty}{\to} 0$ and $\{\psi_j\}_{j\in\N}\subset L^2(\R^d)$ be bounded, then the set $\{w^{\psi_j}_{\epsi_j}\}_{j\in\N} \subset\CC'$
is weak-$*$ compact and every limit point $\mu\in\CC'$ defines a bounded positive Radon measure, called a Wigner measure of $\{\psi_j\}_{j\in\N}$.
\end{proposition}
\begin{proof}
The Calderon-Vaillan\-court theorem (\ref{CV}) implies that $\{w^{\psi_j}_{\epsi_j}\} \subset\CC'$
is bounded. Hence, it is weak-$*$ compact.
By (\ref{Duality}) and the semiclassical sharp G\aa rding inequality, e.g.\ Theorem 7.12 in \cite{DiSj},
it follows that for each $a\geq 0$ there is some  $C<\infty$
such that
\[
\langle  w^{\psi}_\epsi,\,a \rangle_{\CC',\CC}\geq -C\,\epsi\, \|\psi\|^2\qquad\mbox{for all}\,\psi\in L^2(\R^d)\,.
\]
This implies the positivity of all limit points in $\CC'$, which therefore define measures.

Let $\mu\in\CC'$ be such a limit point with,
 after possible extraction of a subsequence,
$w^{\psi_j}_{\epsi_j}\stackrel{*}{\rightharpoonup}\mu$. From (\ref{Duality}) it follows that
\[
\langle w^{\psi}_\epsi,\, 1 \rangle_{\CC',\CC} = \|\psi\|^2_{L^2(\R^d)}\qquad\mbox{for all}\,\psi\in L^2(\R^d)\,,
\]
and thus,
\begin{eqnarray*}
\mu(\R^{2d}) &=& \sup\{\mu(K): K\subset\R^{2d}\,\mbox{compact}\} \\&\leq&\langle \,\mu,\,1\rangle_{\CC',\CC} =\lim_{j\to\infty}\, \langle w^{\psi_j }_{\epsi_j},\, 1  \rangle_{\CC',\CC}=\lim_{j\to\infty}\,\|\psi_j \|^2_{L^2(\R^d)}\,.
\end{eqnarray*}
Hence, $\mu$ is bounded.
\end{proof}

However, not all limit points are physically sensible. For example, the bounded sequence
$\psi_j(x) := \psi_0(x-j)\in L^2(\R)$ has a limit point in $\CC'$, some  Banach-limit type functional,
but the corresponding measure  is   zero. More generally, there are many continuous linear functionals on $\CC$
which are zero on the (non dense) subset $C^\infty_0(\R^{2d})$.

\begin{definition}
A sequence $\{\psi_j\}_{j\in\N}$ remains {\em localized in phase space} (with respect to $\{\epsi_j\}_{j\in\N}$), if it is compact at infinity, i.e.\
\[
\lim_{n\to\infty} \limsup_{j\to \infty} \int_{|x|\geq n}\D x\,\,|\psi_j(x)|^2 = 0  \,,
\]
and $\epsi$-oscillatory, i.e.\
\[
\lim_{n\to\infty} \limsup_{j\to \infty} \frac{1}{\epsi_j^d} \int_{|p|\geq n}\D p\,\,|\widehat{\psi_j}(p/\epsi_j)|^2 = 0  \,.
\]
 \er
\end{definition}

\begin{proposition}
Let $w^{\psi_j }_{\epsi_j}\stackrel{*}{\rightharpoonup}\mu$ in $\CC'$ with $\{\psi_j \}_{j\in\N}\subset L^2(\R^d)$ bounded
and localized in phase space,
then $\mu$ has total mass
\be\label{totalmass}
\mu(\R^{2d}) = \lim_{j\to\infty}\,\|\psi_j \|^2_{L^2(\R^d)}\,,
\ee
and its marginals are given through the weak limits (in $\mathcal{M}^+_{\rm b}$) of the quantum mechanical distributions,
i.e.\ for all $a\in C^0_{\rm b}(\R^d)$ one has
\begin{eqnarray}\label{qmar}
   \int\mu(\D q,\D p)\,a(q)  &=& \lim_{j\to\infty}\,\int\D q\,|\psi_j (q)|^2\,a(q)\,,\\
\int\mu(\D q,\D p)\,a(p)  &=& \lim_{j\to\infty}\,\epsi^{-d}_j\,\int\D p\,|\widehat{\psi_j}(p/\epsi_j)|^2\,a(p)\,.\label{pmar}\nonumber
\end{eqnarray}
\end{proposition}

\begin{proof}
We start with the position marginal (\ref{qmar}). Let $a\in C^\infty_{\rm b}(\R^d)$ and let $\{a_n\}_{n\in\N}\subset C^\infty_0(\R^d)$
and $\{\chi_n\}_{n\in\N}\subset C^\infty_0(\R^d)$ satisfy $a_n(q)=a(q)$ and $\chi_n(p)=1$ for $|q|\leq n$ resp.\ $|p|\leq n$.
Then, by dominated convergence,
\begin{eqnarray*}
 \int\mu(\D q,\D p)\,a(q) &=& \lim_{n\to\infty} \int\mu(\D q,\D p)\,a_n(q)\chi_n(p) \\
 &=& \lim_{n\to\infty}\langle \mu,\,a_n\chi_n\rangle_{\CC',\CC} = \lim_{n\to\infty}\lim_{j\to\infty}
 \langle w^{\psi_j}_{\epsi_j},\,a_n\chi_n\rangle_{\CC',\CC}\\
 &=& \lim_{j\to\infty}\,\int\D q\,|\psi_j (q)|^2\,a(q) + R\,,
\end{eqnarray*}
where
\begin{eqnarray*}
|R| &\leq& \lim_{n\to\infty}\lim_{j\to\infty}
 |\langle w^{\psi_j}_{\epsi_j},\,(a-a_n\chi_n)\rangle_{\CC',\CC}|\\&\leq&
 \lim_{n\to\infty}\lim_{j\to\infty}\left(
 |\langle \psi_j,\,(\widehat a-\widehat{a\chi_n})\,\psi_j\rangle|  +
 |\langle \psi_j,\,(\widehat{a\chi_n}-\widehat{a_n\chi_n})\,\psi_j\rangle|\right)
 \\&=&
 \lim_{n\to\infty}\lim_{j\to\infty}\left(
 |\langle \psi_j,\,(\widehat a-\widehat{a}\widehat{\chi_n})\,\psi_j\rangle|  +
 |\langle \psi_j,\,(\widehat{a}\widehat{\chi_n}-\widehat{a_n}\widehat{\chi_n})\,\psi_j\rangle|\right)\\&\leq&
 \lim_{n\to\infty}\lim_{j\to\infty}\left(
  \|\widehat a\psi_j\|\,\| (1-\widehat \chi_n) \psi_j\| +
  \|(\widehat a-\widehat a_n)\psi_j\|\,\|\widehat \chi_n\,\psi_j\|\right)\\&=& 0\,.
\end{eqnarray*}
For the last equality we used that $\{\psi_j\}$ is localized in phase space. In order to prove
(\ref{qmar}) also for $a\in C^0_{\rm b}$ note that we just proved that the right hand side of (\ref{qmar})
defines a measure. Hence, the result follows again by dominated convergence.
 The statements about the momentum marginal and the total mass follow analogously.
\end{proof}

We now turn to the propagation of Wigner measures.
As remarked in the introduction, a popular approach to the semiclassical limit of (\ref{scaled_dynamics})
is to determine the resulting transport equation for the Wigner measure associated with an $\epsi$-dependent initial condition

\begin{corollary}\label{MeasureCor}
Let $E_n$ be an isolated, non-degenerate Bloch band. Let $\mu_0$ be the   Wigner
measure of a bounded sequence $\{\psi_{0,j}\}$ with $\psi_{0,j} \in \Pi^{\epsi_j}_n L^2(\R^d)$, i.e.\ $w^{\psi_{0,j} }_{\epsi_j} \stackrel{*}{\rightharpoonup} \mu_0\in \CC'$.

Then the Wigner function $w_{\epsi_j}^{\psi_{t,j} }$ of the the time-evolved sequence
\[
\psi_{t,j}  := \E^{-\I H^{\epsi_j} t/{\epsi_j}}\psi_{0,j}
\]
has the  weak-$*$ limit
$\mu_t\in \CC_{\rm per}'$   given through
\be\label{transport}
\mu_t = \mu_0 \circ \overline \Phi^{\,-t}_0\,.
\ee
In particular, $\mu_t$ is a positive bounded measure and solves the transport equation
\[
\dot \mu + \nabla E_n (k-A(r)) \cdot \nabla_r \mu - \Big( \nabla \phi(r) - \partial_l E_n \big(k-A(r)\big) \nabla A_l(r) \Big) \cdot \nabla_k \mu =0
\]
in the distributional sense.
\end{corollary}

Similar results were proved in
\cite{MMP,GMMP,BFPR} for the case of vanishing external potentials $A$ and $\phi$.
For   vanishing magnetic potential $A$ but nonzero electric potential $\phi$ they follow from the
results in \cite{HST} or \cite{BMP}.

\begin{proof}[Proof of Corollary~\ref{MeasureCor}]\,
According to Corollary~\ref{WigCor} we have for $a\in \CC_{\rm per}$ that
\[
 \left|  \left\langle \left( w^{\psi_{t,j} }_{\epsi_j}- w^{\psi_{0,j}}_{\epsi_j}\circ \overline\Phi_0^{\,-t}\right) , a \right\rangle_{\CC',\CC}   \right|\leq \epsi_j\,C\,d_\CC(a)\,\|\psi_{0,j}\|^2\,.
\]
 Taking the limit $j\to\infty$ on both sides yields the existence of the limit $\mu_t$ and at the same time (\ref{transport}). The transport equation for $\mu_t$ follows by taking a time-derivative in (\ref{transport})
and recalling that $\overline \Phi_0^t$ is the Hamiltonian flow of (\ref{Hsc}).
\end{proof}

\noindent{\bf Acknowledgements.} We are grateful to Caroline Lasser for helpful discussions on Wigner measures.
This work was supported by the priority program ``Analysis, Modeling and Simulation of   Multiscale Problems'' of the German Science
Foundation (DFG).

\end{document}